\documentclass[conference]{IEEEtran}
\IEEEoverridecommandlockouts
\usepackage{cite}
\usepackage{hyperref}
\usepackage{amsmath,amssymb,amsfonts}
\usepackage{algorithmic}
\usepackage{graphicx}
\usepackage{textcomp}
\usepackage{xcolor}
\def\BibTeX{{\rm B\kern-.05em{\sc i\kern-.025em b}\kern-.08em
    T\kern-.1667em\lower.7ex\hbox{E}\kern-.125emX}}
\begin{document}

\title{Enhancing Architecture Frameworks by Including Modern Stakeholders and their Views/Viewpoints \\
%{\footnotesize \textsuperscript{*}Note: Sub-titles are not captured in Xplore and
%should not be used}
%\thanks{Identify applicable funding agency here. If none, delete this.}
}

\author{\IEEEauthorblockN{Armin Moin}
\IEEEauthorblockA{\textit{Department of Computer Science} \\
\textit{University of Colorado, Colorado Springs}\\
CO, USA \\
amoin@uccs.edu \textsuperscript{*}Corresponding author}
~\\
\and
\IEEEauthorblockN{Atta Badii}
\IEEEauthorblockA{\textit{Department of Computer Science} \\
\textit{University of Reading}\\
Reading, United Kingdom \\
atta.badii@reading.ac.uk}
~\\
\and
\IEEEauthorblockN{Stephan Günnemann}
\IEEEauthorblockA{\textit{School of Computation, Information, and Technology} \\
\textit{and Munich Data Science Institute (MDSI)}\\
\textit{Technical University of Munich}\\
Munich, Germany \\
s.guennemann@tum.de}
~\\
\and
\IEEEauthorblockN{Moharram Challenger}
\IEEEauthorblockA{\textit{Department of Computer Science} \\
\textit{University of Antwerp}\\
\textit{and Flanders Make}\\
Antwerp, Belgium \\
moharram.challenger@uantwerpen.be}
}

\maketitle

\begin{abstract}
Various architecture frameworks for software, systems, and enterprises have been proposed in the literature. They identified several stakeholders and defined modeling perspectives, architecture viewpoints, and views to frame and address stakeholder concerns. However, the stakeholders with data science and Machine Learning (ML) related concerns, such as data scientists and data engineers, are yet to be included in existing architecture frameworks. Only this way can we envision a holistic system architecture description of an ML-enabled system. Note that the ML component behavior and functionalities are special and should be distinguished from traditional software system behavior and functionalities. The main reason is that the actual functionality should be inferred from data instead of being specified at design time. Additionally, the structural models of ML components, such as ML model architectures, are typically specified using different notations and formalisms from what the Software Engineering (SE) community uses for software structural models. Yet, these two aspects, namely ML and non-ML, are becoming so intertwined that it necessitates an extension of software architecture frameworks and modeling practices toward supporting ML-enabled system architectures. In this paper, we address this gap through an empirical study using an online survey instrument. We surveyed 61 subject matter experts from over 25 organizations in 10 countries.
\end{abstract}

\begin{IEEEkeywords}
architecture frameworks, viewpoints, views, machine learning, modeling, empirical research
\end{IEEEkeywords}

\section{Introduction}\label{sec:introduction}
\textit{Architecture frameworks} provide conventions, principles, and practices for architecture descriptions in a particular application domain or stakeholder community \cite{ISO-IEC-IEEE-42010-2011}. There exist several well-established examples, including The Open Group Architecture Framework (TOGAF) \cite{TOGAF2006, TOGAF2018}, the U.S. Department of Defense Architecture Framework (DoDAF) \cite{DODAF2010}, the Treasury Enterprise Architecture Framework (TEAF) \cite{TEAF2000}, the British Ministry of Defence Architecture Framework (MODAF) \cite{MODAF2012}, the Zachman Framework \cite{Zachman1987}, the “4+1” View Model of Software Architecture \cite{Kruchten1995} and its updated version with the Decision View \cite{Kruchten+2009}, as well as the Reference Model of Open Distributed Processing (RM-ODP) \cite{RM-ODP-ISO-IEC-10746-1, RM-ODP-ISO-IEC-10746-4-1998, RM-ODP-ISO-IEC-10746-2-2009, RM-ODP-ISO-IEC-10746-3-2009}. TOGAF, DoDAF, TEAF, and MODAF were primarily concerned with enterprise architectures. MODAF was replaced by the NATO Architecture Framework (NAF), which in its fourth version (NAFv4) \cite{NAFv4} provided guidance not only on describing enterprise architectures but also system architectures for military and business use. Furthermore, the Zachman Framework, a generic framework for information systems and enterprise architectures, and the “4+1” View Model of Software Architecture were among the early works that offered the foundations for the more recent architecture frameworks. Last but not least, RM-ODP was not merely a reference model but a set of four international standards, including an architecture framework for distributed information processing in heterogeneous environments. Each of the above-mentioned architecture frameworks identified stakeholders, such as end-users, software developers, system integrators, system engineers, business domain experts, executives, and other corporate functions \cite{Kruchten1995, TOGAF2018}. Also, they defined architecture viewpoints and views. Section \ref{sec:background} provides some background on architectural artifacts, such as viewpoints and views.

However, prior work in architecture frameworks did not recognize any stakeholder with ML-related concerns, although ML is increasingly making software and information systems smart, and ML components are assuming a prominent role in many systems and organizations. Consequently, existing architecture frameworks lack any viewpoint or view dedicated to the stakeholders of ML artifacts. Hence, we argue that architecture descriptions of smart (i.e., ML-enabled) software systems, which need to reflect on their ML components and the interactions of the ML components with the non-ML components, cannot be adequately designed using state-of-the-art architecture frameworks. Essentially, ML is a separate field, a sub-discipline of Artificial Intelligence (AI) rather than Software Engineering (SE). Thus, it has its vocabulary, skill set, and know-how, which are different from the ones possessed by typical software developers. Recent research work, such as the interviews conducted by Nahar et al. \cite{Nahar+2022}, stressed the necessity of collaboration between software engineers and other specialists, such as data scientists, for building ML-enabled systems and the associated challenges. In particular, they emphasized the human factors of collaboration, including the need to separate the data science and SE work, as well as to coordinate between them, negotiate and document interfaces and responsibilities, and plan the system operation and evolution. According to them, those human collaboration challenges appeared to be the primary obstacles in developing ML-enabled systems. Additionally, past work mainly focused on ML models, such as the challenges of learning, testing, or serving ML models. They were rarely concerned with the entire system, with many non-ML parts into which the ML model is embedded as a component, which requires coordinating and integrating work from multiple experts or teams \cite{Nahar+2022}. Similarly, Lewis et al. \cite{Lewis+2021} elaborated on the so-called \textit{ML Mismatch} problem that typically occurs in the process of development, integration, deployment, and operation of ML-enabled systems due to incorrect assumptions made about system elements by different stakeholders, such as data scientists, software engineers, and the operations team.

In this paper, we postulate that ML artifacts of smart systems deserve to be treated as having a separate identity, distinguished from software source code and raw data. Moreover, stakeholders should see ML aspects of systems in different ways (i.e., using tailored notations and at various detail levels) such that they can better understand architecture descriptions of ML-enabled systems. In this way, they should become capable of efficient communication with other stakeholders and collaborate to contribute to the system or its architecture description, for example, by rigorous requirement elicitation and tracing. For instance, a data scientist is often in charge of analytics modeling and is primarily concerned with ML performance metrics, such as Accuracy, Precision, and Recall of the ML model when faced with unseen test data. However, a data engineer is responsible for analytics operations, thus ensuring scalable data processing \cite{Pivarski+2016}. By contrast, other stakeholders, such as software engineers, software architects, database engineers, and system engineers, consider other aspects of performance that might be affected by the performance of the ML component, for example, by the delay introduced as a result of the predictions of the ML model, or might be unrelated to the ML component. The same holds for security: software and system engineers might underestimate the potential vulnerability of ML-enabled systems through adversarial attacks on the ML models. Therefore, data scientists and software engineers often have different notions of security. There also exists other challenges, such as the need for versioning ML-artifacts, for example, ML models, or at least their parameters (in the case of parametric ML models) and hyperparameters, monitorability concerns for ML models, as well as new Ethical, Legal, and Social Implications (ELSI), for example, new privacy challenges, or the explainability, fairness, and trustworthiness of ML components.

We argue that existing software system architecture perspectives and viewpoints cannot realize the above-mentioned requirements. As opposed to classic deterministic or stochastic behavioral, functional, or logical models of software systems, ML models rely on inference on data. Apart from certain application scenarios, in most cases, ML models will need to be retrained on new data in the future or even continuously trained in an online learning scenario. For instance, they can become fooled, biased, or compromised through adversarial attacks carried out on the training data that are fed to them. While existing perspectives and viewpoints may provide \textit{model kinds} and architecture views that enable data flow modeling, they essentially lack a systematic approach to inference models and mechanisms, such as mathematical formalisms that can capture the underlying statistical models or Data-Flow Graphs (DFGs), also known as Computational Graphs (CGs) \cite{Abadi+2015}. Moreover, without using the model kinds, formalisms, and notations that are commonly used by ML experts and data scientists, architecture models cannot fully support a fruitful stakeholder discussion and communication, nor can they be used for system documentation or automated code generation. Those are key use cases for system architecture models.

The contribution of this paper is twofold: (i) Identifying ML, data science, and data engineering stakeholders and their concerns; (ii) Proposing modeling perspectives, viewpoints, and views that can frame and address those concerns to enable a sophisticated and thorough architecture description for ML-enabled systems.

The remainder of this paper is structured as follows: Section \ref{sec:background} provides a brief background, whereas Section \ref{sec:sota} reviews the related work in the literature. Further, Section \ref{sec:research-design} elaborates on the research design and methodology. Moreover, Section \ref{sec:enhancing-architecture-frameworks} proposes new stakeholders, viewpoints, and views. Also, Section \ref{sec:threats-to-validity} points out possible threats to validity. Finally, Section \ref{sec:conclusion-future-work} concludes and suggests future work.

\section{Background}\label{sec:background}
\begin{figure}[ht]
	\centering
	\includegraphics[width=0.4\textwidth]{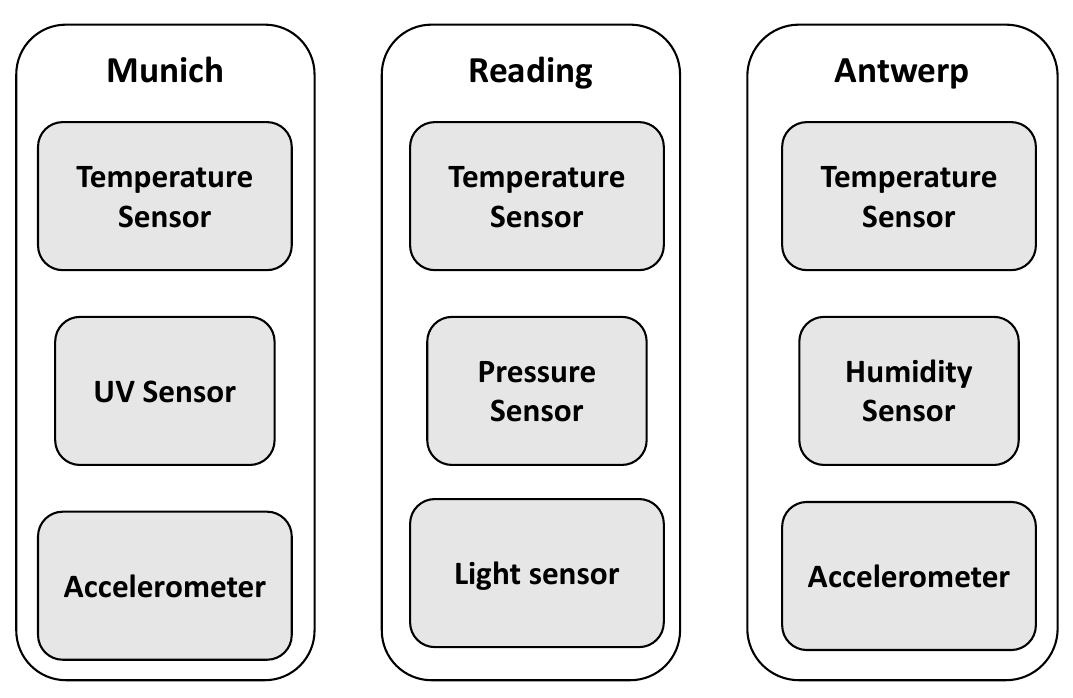}
	\caption{A sample architecture view in TOGAF \cite{TOGAF2018} governed by the viewpoint of Table \ref{tab:TOGAF-sample-viewpoint}}
	\label{fig:TOGAF-sample-view}
\end{figure}

The \textit{ISO/IEC/IEEE 42010:2011} standard for architecture descriptions in systems and software engineering \cite{ISO-IEC-IEEE-42010-2011} defines the \textit{architecture} of a system as fundamental concepts or properties of the \lq{}system in its environment embodied in its elements, relationships, and in the principles of its design and evolution\rq{} \cite{ISO-IEC-IEEE-42010-2011}. An \textit{Architecture Description (AD)} is a work product that expresses an architecture. Every system \textit{stakeholder} has various \textit{concerns} regarding the System under Consideration (SuC), also known as the System of Interest (SoI), in relation to its environment. Concerns arise throughout the life cycle of the system from the system requirements, design choices, implementation, and operations. Performance, reliability, security, privacy, distribution, openness, evolvability, modularity, cost, and regulatory compliance are a number of examples of concerns \cite{ISO-IEC-IEEE-42010-2011, TOGAF2018}. Furthermore, \textit{Separation of Concerns (SoC)}, which is a vital \textit{design principle} in SE, can be applied at different abstraction levels. At a lower level, it is interpreted as the modularity of the software system implementation, with information hiding and encapsulation in modules that have well-defined interfaces. However, at a higher level, it means describing the system architecture from the perspective of different sets of concerns (i.e., different \textit{architecture viewpoints}). This is what an \textit{architecture view} does. It is a work product that expresses \lq{}the architecture of a system from the perspective of specific system concerns\rq{} \cite{ISO-IEC-IEEE-42010-2011}. Additionally, an \textit{architecture viewpoint} is a work product that establishes \lq{}the conventions for the construction, interpretation, and use of architecture views to frame specific system concerns\rq{} \cite{ISO-IEC-IEEE-42010-2011}. In other words, a view is what a stakeholder can see, whereas a viewpoint is where the stakeholder is looking from (i.e., the perspective or template that determines what they should see). A viewpoint is generic and can be stored in a library for reuse. However, a view is specific to the architecture for which it is created \cite{TOGAF2018}. For instance, Table \ref{tab:TOGAF-sample-viewpoint} and Figure \ref{fig:TOGAF-sample-view} illustrate a sample architecture viewpoint and its corresponding architecture view, respectively, based on the TOGAF architecture framework. The view describes the architecture of a Cyber-Physical System (CPS) concerning the physical location of the sensors in the distributed system.

\begin{table}[!ht]
	\caption{A sample architecture viewpoint in TOGAF \cite{TOGAF2018}}
	\begin{center}
		\begin{tabular}{|p{1.75cm}|p{6.5cm}|}
		    \hline
		    \textbf{Element} & \textbf{Description} \\
		    \hline
		    Stakeholders & Chief Technology Officer, system engineer, system integrator \\
		    Concerns & Show the top-level relationships between geographical sites and deployed sensors. \\
		    Modeling technique & Nested boxes diagram. Outer boxes = locations; inner boxes = sensors. The semantics of nesting = sensors deployed at the locations. \\
		    \hline
		\end{tabular}
		\label{tab:TOGAF-sample-viewpoint}
	\end{center}
\end{table}

The “4+1” View Model of Software Architecture \cite{Kruchten1995}, which was proposed in 1995 and updated in 2009 \cite{Kruchten+2009}, considers the following architecture viewpoints\footnote{Note that they used the term view in their work, which corresponded to the notion of viewpoint as defined by ISO/IEC/IEEE 42010:2011 \cite{ISO-IEC-IEEE-42010-2011}.} for software systems: i) The \textbf{logical} viewpoint concentrates on the functionality of the software system. The key stakeholder group whose concerns are framed by this viewpoint is the end-user group. Moreover, the components in the views associated with this viewpoint are often classes. ii) The \textbf{process} viewpoint focuses on the non-functional aspects, such as performance, availability, fault tolerance, integrity, and scalability. This viewpoint is considered primarily for system integrators and system engineers. Furthermore, the components in the views associated with this viewpoint are typically tasks. iii) The \textbf{development} viewpoint is concerned with the software development and management aspects, such as the subsystems and modules' organization, reuse, and portability. The main stakeholder groups here are the programmer group (i.e., software developers) as well as software managers. Also, the components in the views associated with this viewpoint are usually modules and subsystems. iv) The \textbf{physical} viewpoint deals with scalability, performance, and availability on the physical layer, thus considering the network communications, distribution, topology, and the mapping of software onto the hardware. The main stakeholder group here is the system engineer group. Moreover, the components in the views associated with this viewpoint are often nodes. v) The \textbf{scenarios} viewpoint ensures understandability and includes a small set of use case scenarios that can show how the elements of the above-mentioned viewpoints can work together. In fact, this viewpoint is an abstraction of the system requirements. It is worth mentioning that this viewpoint is in some sense redundant from the other ones. This is why the name contains \lq{}+1\rq{}. However, it serves the important purpose of being a driver to discover the architectural elements during the design phase and playing a validation and illustration role after the completion of the design for the test of an initial system prototype. This viewpoint is mainly intended for end-users and software developers. Lastly, the components of the views associated with this viewpoint are steps and scripts. vi) The \textbf{decision} viewpoint: Architects have various design choices and make a lot of design decisions. These decisions must be rigorously documented. The focus of this new viewpoint in the revised version of the original \lq{}4+1\rq{} model is on design decisions.

Further, the aforementioned standard \cite{ISO-IEC-IEEE-42010-2011} defined a \textit{model kind} as a set of conventions for a type of modeling. An architecture viewpoint comprises one or more model kinds. Similarly, an architecture view comprises one or more architecture models. Additionally, each and every architecture viewpoint governs one or more architecture views; also, each and every model kind governs one or more architecture models. For instance, \lq{}data flow diagrams, class diagrams, Petri nets, balance sheets, organization charts, and state transition models\rq{} are model kinds \cite{ISO-IEC-IEEE-42010-2011}. Finally, \textit{architecting} is the \lq{}{process of conceiving, defining, expressing, documenting, communicating, certifying proper implementation of, maintaining, and improving an architecture throughout a system life cycle\rq{} \cite{ISO-IEC-IEEE-42010-2011}. It is typically conducted in the context of a project or an organization.

\section{Related Work}\label{sec:sota}
The “4+1” View Model of Software Architecture, as explained in Section \ref{sec:background}, was an early but leading work concerned with the architecture of software systems. Other architecture frameworks, such as the Zachman Framework \cite{Zachman1987}, which preceded this, or the other ones that succeeded it (see Section \ref{sec:introduction}), had also identified a number of stakeholders, as well as several viewpoints and views. In the following, we refer to DoDAF \cite{DODAF2010} and TOGAF \cite{TOGAF2006, TOGAF2018} whose latest versions were published in 2010 and 2018, respectively, as two representative examples. In particular, we highlight their data-related aspects since those are the most related ones to the target of this study.

DoDAF \cite{DODAF2010} defined several viewpoints including the Data and Information Viewpoint 1 (DIV-1, called \textbf{conceptual data} model), DIV-2 (called \textbf{logical data} model), and DIV-3 (called \textbf{physical data} model). DIV-1 was mainly intended for non-technical stakeholders and showed the high-level data concepts and their relationships. Relationships at this level would be simple (i.e., not attributed). However, DIV-2 bridged the gap between the conceptual and physical levels by introducing the attributes and structural business process (activity) rules that formed the data structure. This viewpoint was intended for system architects and analysts. For example, an Entity-Relationship (ER) diagram or a Class diagram could be deployed as a candidate model kind for this viewpoint. Lastly, DIV-3 provided the physical schema for the data and information and was close to the actual implementation level. This could often be represented by tables, records, and keys in a relational database management system or through objects in an object-oriented data model. The physical data model (i.e., DIV-3) was intended for database engineers, software developers, and system engineers.

Moreover, the TOGAF standard \cite{TOGAF2006, TOGAF2018} is closely related to the Zachman Framework \cite{Zachman1987}\footnote{In fact, there exists a mapping \cite{TOGAF2006} between the TOGAF Architecture Development Method (ADM) and the Zachman Framework.}. In its 2006 version (i.e., v8.1.1), TOGAF \cite{TOGAF2006} adopted the data flow viewpoint\footnote{Note that they used the term view in their work, which corresponded to the notion of viewpoint as defined by ISO/IEC/IEEE 42010:2011 \cite{ISO-IEC-IEEE-42010-2011}.}, which was intended for database engineers, \lq{}concerned with the storage, retrieval, processing, archiving, and security of data\rq{}, thus \lq{}assuring ubiquitous access to high-quality data\rq{} \cite{TOGAF2006}. Further, the latest version of TOGAF (v9.2) from 2018 \cite{TOGAF2018} promoted a set of data-related viewpoints (which they called data \textit{diagrams}): i) The \textbf{conceptual data} viewpoint had the key purpose of depicting the relationships between the critical data entities of an enterprise. This viewpoint was intended for business stakeholders. ER diagrams or simplified UML Class diagrams could be deployed as the model kinds supporting this viewpoint. ii) The \textbf{logical data} viewpoint was intended for application developers and database designers and showed the logical views of the relationships between the critical data entities within an enterprise. iii) The \textbf{data dissemination} viewpoint illustrated the relationship between data entity, business service, and application components. It showed how the logical entities were physically realized by application components. iv) The \textbf{data security} viewpoint intended to depict which actor (i.e., person, organization, or system) has access to which enterprise data. This could be demonstrated in a matrix form or as a mapping. Moreover, it could show compliance with data privacy laws and other applicable regulations. v) The \textbf{data migration} viewpoint showed the flow of data between the source and target applications when implementing a package or packaged service-based solution, for example, if a legacy application was to be replaced. Packages might have their own data model. Thus, data transformation might be necessary. The transformation could include data quality processes, namely data cleansing, matching, merging, and consolidating data from different sources, as well as source-to-target mappings. This viewpoint could be deployed for data auditing and establishing traceability. The detail level of the supporting views could vary. vi) The \textbf{data life cycle} viewpoint enabled managing business data throughout their life cycle from conception until disposal. Each change in the state of data should be shown through the views supporting this viewpoint. The data were considered entities in their own rights, decoupled from business processes and activities. This allowed common data requirements to be identified.

It is evident that none of the architecture frameworks included any stakeholders, perspectives, viewpoints, views, or model kinds that support data scientists and their collaboration and communication with other stakeholders in the course of system architecture design for ML-enabled systems, which is required in Artificial Intelligence (AI) Engineering.

Recently, various works have pointed out the need for upgrading architecture frameworks to support ML-enabled systems. Almost around the same time that we conducted our empirical study (2021), Muccini and Vaidhyanathan \cite{MucciniVaidhyanathan2021} published a related research paper based on their experience in architecting an ML-based software system. They identified the following stakeholders with ML-related concerns: data scientists, ML developers, ethics experts, and data engineers. Moreover, they envisioned the following concerns for them: ML model accuracy, model versioning, data quality, privacy, ethics, framework, and algorithmic choice. Our empirical results confirmed their identified stakeholders and had some overlap with their stated concerns.

Additionally, Lewis et al. \cite{Lewis+2021a} illustrated the role of architects in ML-enabled systems and highlighted some new challenges faced by architects and other stakeholders in devising software architecture of ML-enabled software systems, namely software systems that rely on one or more ML components (including ML models) to provide their capabilities. Their study was also published in 2021 when we conducted our survey. Their work emphasized the importance of bringing the SE and data science communities together to address the concerns of both stakeholder groups holistically. They also agreed with our understanding that ML artifacts must be treated as special software artifacts, and existing principles and practices in architecture frameworks must be updated to address new concerns emerging from these artifacts.

We started from our own experience to identify the gap in the literature and formulate the research questions. However, unlike most of the related work, we conducted an empirical study that referred to domain experts and asked them for their input, positions, and preferences on the matter. Note that Lewis et al. \cite{Lewis+2021} also conducted practitioner interviews as part of their work. They showed that the development and deployment of ML-enabled systems involve three different perspectives, which include data science, software engineering, and operations. If they are misaligned due to incorrect assumptions, they cause ML mismatches, which can result in failed systems. Their study was also published in 2021 when we surveyed our experts.

\section{Research design and methodology}\label{sec:research-design}
While literature review surveys, systematic literature reviews, and meta-analyses are valuable, we lack sufficient qualitative and quantitative empirical studies that offer expert visions and open up new horizons for software engineering research. Therefore, given the new nature of the topic at the intersection of software system architecture frameworks and ML, we decided to carry out such a study with 65 subject matter experts from more than 25 organizations in over 10 countries\footnote{The employment location of each expert at the interview or survey time was taken into account.} (Belgium, Canada, Denmark, France, Germany, Serbia, Switzerland, Turkey, the United Kingdom, and the United States) to devise and validate the proposed framework. 

We postulated the following Research Questions (RQs): RQ1) Who has a stake in ML-enabled software systems and their architecture design? RQ2) What are the preferred formalisms, notations, and, in general, model kinds supporting viewpoints and their corresponding views for each stakeholder group, in particular, for collaborating with those in the data science domain, who often use different model kinds from what the Software Engineering (SE) community uses?

We conducted research in two steps: 1) We interviewed four selected domain experts to gain some insight into the problem domain and validate the interview questions. To this aim, we deployed the \textit{Qualitative Surveys (Interview Studies)} empirical research method \cite{ACM-Empirical-Standard-Qualitative-Survey}. 2) We adopted the quantitative empirical research method, called \textit{Questionnaire Surveys} \cite{ACM-Empirical-Standard-Quantitative-Questionnaire-Surveys} and carried out an online survey with 61 participants.

\paragraph*{Step 1}
We interviewed four experts from the authors' networks from May to June 2021. The interviewee selection was based on convenience sampling (i.e., not random). The interviewee profiles were as described in Table \ref{tab:interviewee_first_round_demographics}. The numbers in parentheses denote the frequencies. The interviews were semi-structured with open questions.

\begin{table}[!ht]
	\caption{Summary of the demographic information of the interview partners}
	\begin{center}
		\begin{tabular}{|p{2.5cm}|p{5.5cm}|}
		    \hline
		    \textbf{Type} & \textbf{Breakdown} \\
		    \hline
            Sex or gender (4) & female (1), male (3), other (0), no answer (0) \\
            Age group (4) & below 18 (0), 18-24 (0), 25-39 (4), 40-60 (0), 60 plus (0), no answer (0) \\
            Highest degree (4) & Bachelor's (1), Master's (1), Ph.D. (2), No academic degree (0), Other (0), no answer (0) \\
            Field of expertise (4) & ML \& SSE (1), Model-Driven Engineering (2), general SSE (1) \\
            Job or occupation (4) & researcher (1), senior software engineer (1), sales software engineer (1), data scientist (1) \\
		    \hline
		\end{tabular}
		\label{tab:interviewee_first_round_demographics}
	\end{center}
\end{table}

\paragraph*{Step 2}
We carried out a survey study from July to September 2021 to answer RQ1 and RQ2 above. The survey questionnaire was offered through a link (URL) as an online\footnote{We used the open-source LimeSurvey software \cite{limesurvey} on our own server.}, self-administered survey, with the option of fully anonymous participation in the study. However, we collected the IP addresses to prevent any possible redundant participation. The questionnaire had four sections and a total of 25 questions. The order of the questions in each section and the order of the choices in the case of multiple-choice questions would be set on a random basis for each participant. We had a total of 121 participants, out of which 60 participants did not answer the questionnaire at all. Therefore, we took the results of the remaining 61 participants into account. The selection process for the invitation of the subject matter experts to participate in this study was again based on convenience sampling (i.e., not random) through the authors' networks, for example, by direct invitation of peers via email and sharing the URL on LinkedIn. Part of the participants' demographic information is summarized in Table \ref{tab:survey_participants_demographics} (the numbers in parentheses denote the frequencies). The average participation time was 14 minutes, whereas the median was 11 minutes. All questions were optional to answer except for the two questions regarding the consent of the survey participants with respect to their anonymity and receiving a pre-print of the study in the future.

\begin{table}[!ht]
	\caption{Summary of the survey participants' demographic information}
	\begin{center}
		\begin{tabular}{|p{1.5cm}|p{6.5cm}|}
		    \hline
		    \textbf{Type} & \textbf{Breakdown} \\
		    \hline
            Sex or gender (61) & female (5), male (31), other (0), no answer (25) \\
            Age group (61) & below 18 (0), 18-24 (2), 25-39 (30), 40-60 (6), 60 plus (1), no answer (22) \\
            Highest degree (61) & Bachelor's (3), Master's (21), Ph.D. (17), No academic degree (0), Other (0), no answer (20) \\
            DEA and ML expertise (61) & beginner (6), medium level (18), expert (16), no self estimation (i.e., don't know) (2), no answer (19) \\
            SE expertise (61) & beginner (7), medium level (14), expert (18), no self estimation (i.e., don't know) (2), no answer (20) \\
            Job or occupation (61) & data scientist \& ML engineer (12), data engineer (1), software engineer (6), software architect (4), system engineer (4), data science \& ML researcher (3), SE researcher (2), CS student (1), software community manager (1), CTO (1), software tester (1), no answer (25) \\
		    \hline
		\end{tabular}
		\label{tab:survey_participants_demographics}
	\end{center}
\end{table}

\section{Enhancing architecture frameworks}\label{sec:enhancing-architecture-frameworks}
Existing software system architecture frameworks fall short of supporting a holistic architecture description for ML-enabled systems. Some members of the Software Engineering (SE) community may argue that the ML-related functionality might be specified using existing functional, behavioral, or logical model kinds, and perhaps the structure of ML models might be specified using existing structural model kinds. However, looking at the reality and state of practice in ML, data science, and data engineering, as well as the expectations of the subject matter experts in those fields, it is evident that this wish is simply not feasible. ML components are indeed software artifacts, but they are very special ones. As also stated by Lewis et al. \cite{Lewis+2021a}, besides the data-dependent behavior of ML models, the necessity of detecting and responding to drift over time and timely capture of ground truth to inform retraining are among some of the unique characteristics of ML components that pose new challenges to software architects of ML-enabled systems and bring new quality attribute concerns (e.g., with respect to monitorability). Hence, we have to extend our software architecture frameworks if we want our software system architecture descriptions to include this vital aspect of ML-enabled systems, which is affecting the software systems' behaviors and even structures in an unprecedented manner.

In this section, we enhance architecture frameworks by proposing new stakeholders, viewpoints, and views, as explained below.

\subsection{Identified stakeholders}\label{subsec:identified-stakeholders}
In the following, we present the list of identified stakeholder groups for modern systems, software, and enterprises. In particular, we concentrate on the recently emerged ones who may have ML-related concerns, such as data scientists and data engineers.

The following stakeholder groups have already been considered in prior architecture frameworks: 1) end-users, 2) business stakeholders, 3) database designers and engineers, 4) software architects and engineers (i.e., developers), 5) system designers, engineers, and integrators. Additionally, we believe that 6) network engineers and 7) security experts should be distinguished from system engineers and software engineers, respectively, given the sophisticated level of knowledge and skills that are required for designing and managing secured, pervasive technologies of modern systems and organizations. Furthermore, we noticed the stakeholder groups below during the interviews: 8) safety and regulatory compliance engineers, 9) data protection (privacy) officers, and 10) ethics committees or boards. Also, the online survey participants pointed out the following stakeholder groups: 11) quality assurance (test) engineers and 12) maintenance managers. Last but not least, we propose counting 13) data scientists (including ML engineers) and 14) data engineers among the stakeholders of modern systems, software, and enterprises, which often contain ML components or ML-enabled (sub-)systems. The proposed stakeholder groups were validated through the interviews and the online survey.

Data scientists are responsible for analytics modeling. In fact, the task of developing methods for building efficient Data Analytics (DA) models (including ML models) to enable systems that can analyze data and \textit{learn} from data lies at the core of data science. However, there is also a need for technologies regarding the deployment of DA models in products, services, and operational systems. This part is known as analytics operations. Data engineers are typically concerned with this part, which is also called Data Engineering (DE) \cite{Pivarski+2016}. Together, DA (i.e., data science) and DE are called Data Engineering and Analytics (DEA) \cite{TUM-DEA} or Data Science and Engineering (DSE) \cite{Raj+2019}. The typical workflow comprises analytics modeling (e.g., training ML models) by data scientists, followed by the integration and deployment of the data analytics artifacts (e.g., the trained ML models) in the data analytics and ML components, as well as in the larger systems (i.e., ML-enabled systems) by data engineers in collaboration with software engineers, database engineers, system engineers, etc. Afterward, the system should be handed over to the operations team \cite{Lewis+2021}.

Lastly, it is clear that DA models should not be confused with data models, datasets, or data instances. For instance, a DA (e.g., ML) model can be a Probabilistic Graphical Model (PGM), an Artificial Neural Network (ANN), or a Hidden Markov Model (HMM), which may enable inference on data. By contrast, raw data in datasets or data streams have a different nature. For example, one can use raw data to \textit{train} an ML model. Also, a data model refers to an abstract model of the entities and their relationships in a database. Therefore, data models must be distinguished from DA models. Similarly, data scientists and data engineers should not be confused with database engineers.

\subsection{Proposed viewpoints and views}\label{subsec:proposed-viewpoints-views}
We propose two new architecture viewpoint categories to frame the concerns of data scientists and data engineers in the architecture frameworks of systems, software, and enterprises. We call the new viewpoint categories analytics modeling (alternatively DA or data science) and analytics operations (or DE), respectively \cite{Pivarski+2016}. Moreover, we propose adopting and adapting existing notations and model kinds to realize corresponding views for the new viewpoints, as well as new views for the viewpoints of other stakeholders communicating and collaborating with data scientists and data engineers.

According to our own knowledge about the state of practice in the field of DEA (or DSE), we envisioned the mathematical notations commonly used in DA and ML, as well as the graphical notation of PGMs \cite{Bishop2006}, and the Data-Flow Graphs (DFGs), also known as Computational Graphs (CGs) \cite{Abadi+2015} for ANNs, as potentially adequate candidates for providing the necessary model kinds that should serve the architecture views required for supporting the proposed new architecture viewpoints. In addition, we assumed that existing UML diagrams could be adopted for adapting the viewpoints of other stakeholders in order to enable their appropriate communication and collaboration with data scientists and data engineers. The empirical study through the survey questionnaire confirmed these assumptions and helped us devise new viewpoints and views using existing notations and model kinds, as illustrated in Table \ref{tab:viewpoints-views}. The list of notations and model kinds in each row of the table is ordered based on the opinions of the survey participants concerning the suitability of each option for the specific purpose. Tables \ref{tab:data-scientist-1} and \ref{tab:data-scientist-2} show this from the viewpoint of data scientists, whereas Tables \ref{tab:data-engineer-1} and \ref{tab:data-engineer-2} illustrate this from the viewpoint of data engineers. In fact, we asked the same question from each stakeholder group in two different ways: (i) What is the best model kind and notation for their collaboration? (ii) What is the most suitable model kind and notation for describing the system architecture from their viewpoint?

\begin{table*}[!ht]
	\caption{Notations and model kinds supporting the viewpoints and views of ML-enabled architecture frameworks for framing and addressing stakeholder concerns with respect to ML}
	\begin{center}
		\begin{tabular}{|p{6cm}|p{10cm}|}
		    \hline
		    \textbf{Stakeholders communicating or collaborating} & \textbf{Notations and model kinds supporting the viewpoints and views} \\
		    \hline
            Data scientists with their peers & i) mathematical notations, ii) charts, diagrams, or plots, iii) DFGs or CGs \cite{Abadi+2015}, iv) PGMs \cite{Bishop2006}, v) data flow diagrams (or UML activity diagrams showing the flow of data rather than the flow of control), for example, for the data analytics pipeline \\
            \hline
            Data engineers with their peers & i) data flow diagrams, ii) UML class diagrams, iii) DFGs or CGs \cite{Abadi+2015}, iv) Entity-Relationship (ER) diagrams, v) mathematical notations \\ 
            \hline
            End-users with data scientists and engineers & i) text documents, ii) charts, diagrams, or plots, iii) tables or matrices, iv) data flow diagrams, v) UML use case diagrams \\
            \hline
            Business stakeholders with data scientists and engineers & i) charts, diagrams, or plots, ii) text documents, iii) tables or matrices, iv) UML use case diagrams \\
            \hline
            Database designers and engineers with data scientists and engineers & i) ER diagrams, ii) UML class diagrams, iii) data flow diagrams, iv) UML use case diagrams, v) tables or matrices \\
            \hline
            Software architects and engineers with data scientists and engineers & i) UML class diagrams, ii) data flow diagrams, iii) UML use case diagrams, iv) ER diagrams \\
            \hline
            System designers, engineers, integrators, and network engineers with data scientists and engineers & i) UML deployment diagrams, ii) data flow diagrams, iii) DFGs or CGs \cite{Abadi+2015} augmented with physical (i.e., deployment) information, iv) UML class diagrams \\
            \hline
            Security experts with data scientists and engineers & i) data flow diagrams, ii) UML deployment diagrams, iii) DFGs or CGs \cite{Abadi+2015} augmented with physical (i.e., deployment) information, iv) ER diagrams, v) mathematical notations, vi) UML class diagrams \\
            \hline
            Safety and regulatory compliance engineers, data protection (privacy) officers, and ethics committees or boards with data scientists and engineers & i) text documents, ii) data flow diagrams, iii) ER diagrams, iv) DFGs or CGs \cite{Abadi+2015} augmented with physical (i.e., deployment) information, v) tables or matrices, vi) UML deployment diagrams \\
            \hline
		\end{tabular}
		\label{tab:viewpoints-views}
	\end{center}
\end{table*}

\begin{table}[!ht]
	\caption{Notations and model kinds for the collaboration of data scientists with their peers}
	\begin{center}
		\begin{tabular}{|p{6cm}|p{1cm}|}
        \hline
        \textbf{Notation/model kind} & \textbf{Rank based on the votes} \\
		\hline
        Mathematical notation showing the mathematical model (e.g., probability distributions, as well as objective or loss function) & 1st \\
        \hline
        Charts, diagrams, or plots (e.g., histograms, pie charts, and scatter plots) & 2nd \\
        \hline
        Computational Graphs (CG), also known as Data-Flow Graphs (DFG) & 3rd \\
        \hline
        Probabilistic Graphical Models (PGMs) & 4th \\
        \hline
        Data flow diagrams (e.g., using the UML Activity diagram notation) showing the upstream and downstream components in the pipeline & 5th \\
        \hline
        Topic maps, knowledge graphs (e.g., RDF graphs), or Ontologies & 6th \\
        \hline
        UML class diagrams & 7th \\
        \hline
        Entity-Relationship (ER) diagrams & 8th \\
        \hline
		\end{tabular}
		\label{tab:data-scientist-1}
	\end{center}
\end{table}    

\begin{table}[!ht]
	\caption{Notations and model kinds for the software system architecture description from the viewpoint of data scientists}
	\begin{center}
		\begin{tabular}{|p{6cm}|p{1cm}|}
        \hline
        \textbf{Notation/model kind} & \textbf{Rank based on the votes} \\
		\hline
        The workflow/pipeline for data analytics and machine learning & 1st \\
        \hline
        The underlying  mathematical model & 2nd \\
        \hline
        The computational operations and the flow of data among them & 3rd \\
        \hline
        The problem/use case domain concepts and their relationships & 4th \\
        \hline
        The data visualization (e.g., in the case of online learning/stream processing) & 5th \\
        \hline
        The processes and/or components and the flow of data among them & 6th \\
        \hline
        The processes and/or components and the flow of control among them & 7th \\
        \hline
		\end{tabular}
		\label{tab:data-scientist-2}
	\end{center}
\end{table}

\begin{table}[!ht]
	\caption{Notations and model kinds for the collaboration of data engineers with their peers}
	\begin{center}
		\begin{tabular}{|p{6cm}|p{1cm}|}
        \hline
        \textbf{Notation/model kind} & \textbf{Rank based on the votes} \\
		\hline
        Data flow diagrams (e.g., using the UML Activity diagram notation) showing the upstream and downstream components in the pipeline & 1st \\
        \hline
        UML class diagrams & 2nd \\
        \hline
        Computational Graphs (CG), also known as Data-Flow Graphs (DFG) & 3rd \\
        \hline        
        Mathematical notation showing the mathematical model (e.g., probability distributions, as well as objective or loss function) & 4th \\
        \hline
        Entity-Relationship (ER) diagrams & 5th \\
        \hline
        Charts, diagrams, or plots (e.g., histograms, pie charts, and scatter plots) & 6th \\
        \hline
        Topic maps, knowledge graphs (e.g., RDF graphs), or Ontologies & 7th \\
        \hline
        Probabilistic Graphical Models (PGMs) & 8th \\
        \hline
        \end{tabular}
		\label{tab:data-engineer-1}
	\end{center}
\end{table}

\begin{table}[!ht]
	\caption{Notations and model kinds for the software system architecture description from the viewpoint of data engineers}
	\begin{center}
		\begin{tabular}{|p{6cm}|p{1cm}|}
        \hline
        \textbf{Notation/model kind} & \textbf{Rank based on the votes} \\
		\hline
        The processes and/or components and the flow of data among them & 1st \\
        \hline
        The computational operations and the flow of data among them & 2nd \\
        \hline
        The processes and/or components and the flow of control among them & 3rd \\
        \hline
        The workflow/pipeline for data analytics and machine learning & 4th \\
        \hline
        The problem/use case domain concepts and their relationships & 5th \\
        \hline
        The data visualization (e.g., in the case of online learning/stream processing) & 6th \\
        \hline
        The underlying  mathematical model & 7th \\
        \hline
        \end{tabular}
		\label{tab:data-engineer-2}
	\end{center}
\end{table}

\section{Threats to validity}\label{sec:threats-to-validity}
Notwithstanding the limitations in this study arising from the difficulty of ensuring large-scale participation and the concomitant constraints with respect to the sampling choices that had to be made, it is evident that in achieving the elicitation of the opinions and insights of a relatively significant number of expert practitioners, the study has formalized a useful addition to the existing software architecture frameworks to support ML-enabled systems. Below, we point out a number of potential internal and external threats to the validity of this study.

First, not all interview discussion topics and questions were the same. We deliberately matched the topics and questions to the experts' backgrounds and fields of expertise. Also, the expert interviews had various durations (30-60 minutes). The different settings might have affected the results achieved through the interviews.

Second, the sampled participant distribution may not be representative enough to make generalizations with certainty. Furthermore, we neither deployed a randomized method to select the study participants nor could we manage to have all stakeholder groups, roles, jobs, disciplines, or underrepresented populations in this field in academia and in the industry well represented here. For instance, some of the chosen experts had served as the advisors, advisees, or colleagues of some other participants in the study.

Last but not least, similar to other quantitative studies, there could be some construct validity concerns, for example, regarding the use of indicators to measure concepts that were not directly measurable, the choice of metrics, and the assumption on the validity of the interviews and the survey questionnaire, as well as the opinions of the participants.

\section{Conclusion and future work}\label{sec:conclusion-future-work}
In this paper, we have enhanced architecture frameworks to address ML-enabled software systems. We have identified the stakeholders who might have concerns with respect to the ML aspects, namely data scientists and data engineers. Moreover, we have proposed new architecture viewpoint categories (i.e., analytics modeling and analytics operations) as well as model kinds to support these viewpoints and their corresponding views.

The results of this study are expected to improve the effectiveness and increase the efficiency of software development projects involving ML-enabled systems, which require the communication and collaboration of various stakeholder groups with different backgrounds, concerns, metrics, and vocabularies from different domains, organizations, and regions. This expectation is based on the nature of the study, which reflects on the expert knowledge and insight as well as the state of practice in the respective fields.

One limitation of the present study was the absence of particular groups of stakeholders, such as the operations team (see \cite{Lewis+2021}), the maintenance managers, and quality assurance engineers. The two latter items were suggested by our online survey participants. In the future, these and other stakeholder groups can be studied. Also, the enhancement of architecture frameworks for other sub-disciplines of Artificial Intelligence (AI) beyond ML should be helpful.

\section*{Acknowledgments}
The authors would like to thank the following interview and survey participants, as well as the anonymous ones, for dedicating their time to this study. Further, they are grateful for the valuable feedback of the anonymous reviewers and the editors, which improved the quality of this paper. Here is the list of names of the interview partners and survey participants, who opted to share their names, with no specific order: Sebastian Eder (Qualicen GmbH, Germany), Pragya Kirti Gupta (fortiss GmbH, Germany), Bertrand	Charpentier (Technical University of Munich, Germany), Nicholas	Gao (Technical University of Munich, Germany), Günter Neumann (DFKI GmbH, German Research Center for AI, Germany), Shreya Desai (University of Reading, UK), Alper Kanak (ERARGE Co. Ltd., Turkey), Sander Vanhove, Cláudio Gomes (Aarhus University, Denmark), Hussein Marah (University of Antwerp, Belgium), Raheleh Eslampanah (University of Antwerp, Belgium), Onur Kilincceker (University of Paderborn, Germany and Mugla Sitki Kocman University, Turkey), Dušan	Savić (Belgrade University, Serbia), and Ivan Ruchkin (University of Pennsylvania, USA).

\bibliographystyle{IEEEtran}
\bibliography{refs.bib}
\end{document}